\documentclass[nopubldata]{lpor2014}
\usepackage[latin1]{inputenc}
\usepackage{graphicx}
\usepackage{dcolumn}
\usepackage{hyperref}
\usepackage{amsmath}
\usepackage{color}

\hypersetup{%
  colorlinks,
  plainpages=false,
  pdfpagelabels,
  breaklinks,
  pdfview=Fit,
  bookmarksopen,
  bookmarksnumbered,
  linkcolor=black,
  anchorcolor=black,
  citecolor=black,
  filecolor=black,
  urlcolor=LPRred
  }%
\graphicspath{{./figs/}}
\setpages[]{}
\setvolume[0]{0}
\setyear{2011}%
\setdoi{201100000}%
\oddheadlogotext{}{Guide}

\begin{document}

\titlefigure{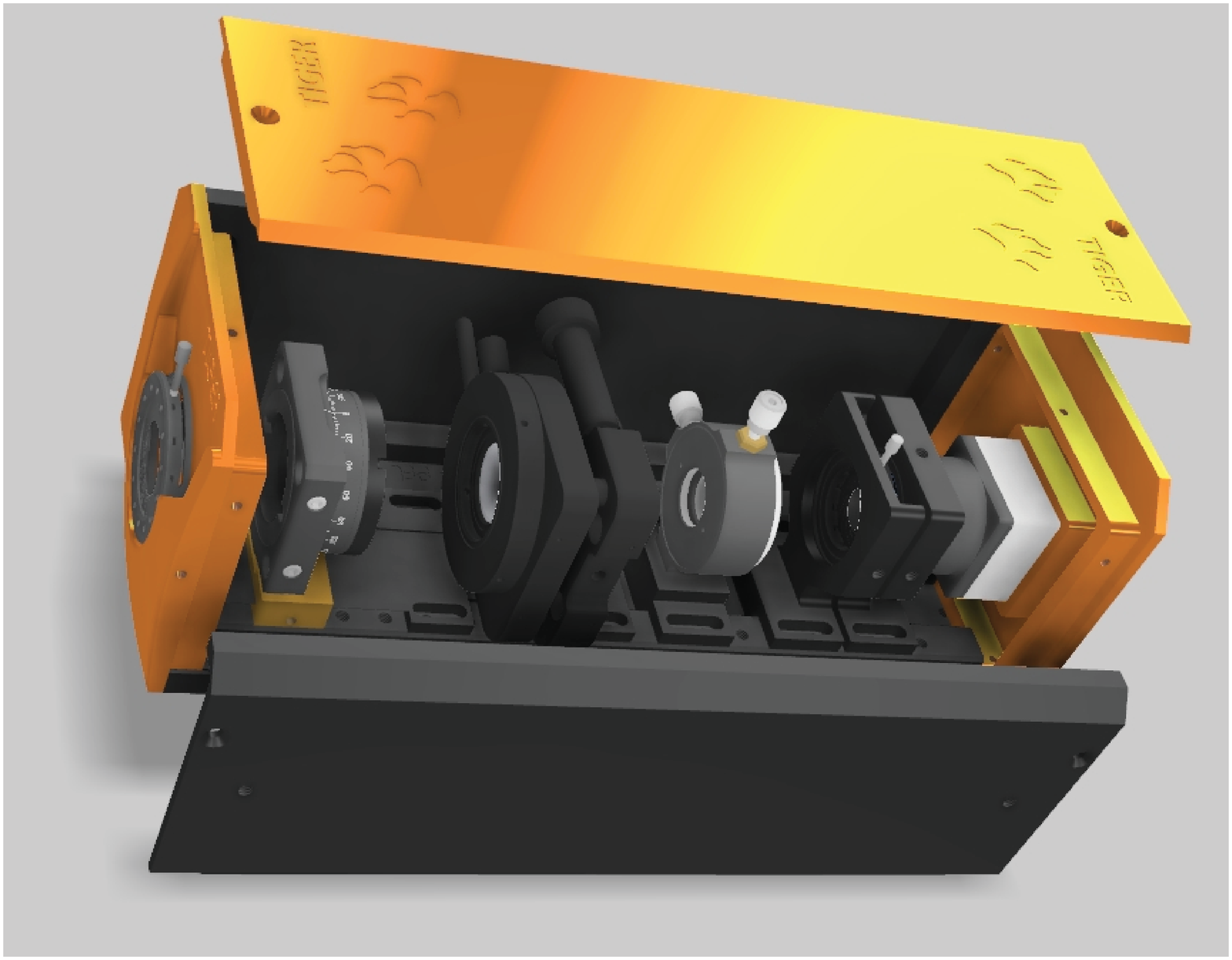}
\abstract{We present a novel self-referenced method for the complete temporal characterization (phase and amplitude) of ultrashort optical laser pulses. The technique, called TIme-Gated Electric field Reconstruction (TIGER), measures a second-order nonlinear signal (namely, second harmonic generation or two-photon absorption) produced by four time-delayed replicas of the input pulse. The delays are spatially encoded in the beam profile using a four-faced pyramid-like optical element. The presented technique enables single shot measurement and does not require any spectral measurements, in contrast with well-known self-referenced characterization methods. Depending on the chosen geometry, the recorded TIGER signal can be either interferometric (i.e., carrier frequency resolved) or intensimetric. This paper describes the principle operation of the device together with a detailed theoretical analysis. TIGER measurements of various laser pulse shapes with their reconstructions are then presented demonstrating the relevance of this original approach. }
\title{TIGER: TIme-Gated Electric field Reconstruction}
\author{F. Billard, R. Sharma, E. Hertz, O. Faucher, and P. B\'ejot$^{*}$}%
\authorrunning{P. B\'ejot}
\mail{\email{pierre.bejot@u-bourgogne.fr}}

\institute{Laboratoire Interdisciplinaire Carnot de Bourgogne, UMR 6303 CNRS/Universit\'e Bourgogne Franche-Comt\'e, 21078 Dijon, France}

\maketitle

\section{Introduction}
The advent of ultrashort laser sources several decades ago has opened a real breakthrough for observing ultrafast phenomena taking place at femtosecond and picosecond timescales and also for studying the behavior of matter submitted to very intense electromagnetic fields. The production of such ultrashort optical events, the shortest timescale ever produced at this time, has immediately raised the question of their measurements \cite{1,2,3}. Indeed, for many applications, the precise knowledge of the time-dependent phase and amplitude of the ultrashort laser pulse is of prime importance. Many self-referenced techniques, with varying experimental complexity and limitations, aiming at characterizing ultrashort laser pulses, have thus been developed. Spectral Phase Interferometry for Direct Electric field Reconstruction (SPIDER) \cite{3,10,11} [resp. Self-Referenced Spectral Interferometry (SRSI) \cite{4}] technique is based on the measurement of the spectral interferences between the pulse to be characterized and a replica that is both spectrally and temporally shifted (resp. a nonlinearly filtered replica). Frequency-Resolved Optical Gating (FROG) \cite{5}, in all its variants, is based on a spectrally resolved measurement of a nonlinear signal produced by two-time delayed replicas. Chirp-scan \cite{6} [resp. Multiphoton Intrapulse Interference Phase Scan (MIIPS) \cite{7}] technique relies on recording the spectrum of a nonlinear signal as a function of a linear frequency chirp [resp. sinusoidal frequency phase] applied to the input pulse. To the best of our knowledge, all the aforementioned techniques are based on spectrally-resolved measurements. On the contrary, the presented Time-Gated Electric field Reconstruction (TIGER) technique is able to reconstruct the spectral characteristics (phase and amplitude) of a pulse from a pure time-domain signal, without the need of direct spectral measurements. The retrieval of spectral information by performing only measurements in the time domain has already been used in various branches of ultrafast optics. For instance, it is well known that recording the linear autocorrelation of a pulse (i.e., the signal intensity produced by two time-delayed replica as the function of the delay) is sufficient to recover the pulse spectrum by a simple Fourier transform, without any need of a spectrometer. Another example of pure time-domain measurements is two-dimensional nonlinear spectroscopy \cite{8,9}, which uses a sequence of more-than-two delayed pulses for probing the nonlinear response of the system under investigation. By Fourier transform, the full spectral characteristics of the system can then be retrieved. The method proposed in this paper can be viewed as a direct application of two-dimensional nonlinear spectroscopy. However, while two-dimensional nonlinear spectroscopy is intended for the study of a sample provided that the laser pulse is known, the TIGER assumes a known (instantaneous) nonlinear response of the material while the pulse characteristics has to be determined. From the nonlinear signal produced by four time-delayed replicas, the TIGER enables a complete characterization (phase and amplitude) of ultrashort pulses by using a retrieval algorithm. The paper is organized as follows. In section \ref{section1}, the principle operation of the TIGER will be developed. First, the way by which four time-delayed pulses are generated will be explained. It will be shown, by geometrical optics calculations, that the use of a four-faced pyramid-like optical element allows to perform single shot measurements while making the device extremely compact. Then, the nonlinear signal produced by the four-time delayed replicas will be derived for both two-photon absorption and second harmonic generation mechanisms. In section \ref{section2}, experimental results will be presented for two distinct configurations. First, results recorded using two-photon absorption in a silicon camera will be detailed. Such a measurement, providing an interferometric signal, enables the characterization of infrared pulses between 1.4 and 2.4\,$\mu$m. Then, intensimetric TIGER measurements, obtained by using second harmonic generation from a Ti:sapphire (Ti:Sa) femtosecond laser, will be discussed. The experimental reconstructions show that the two configurations are very reliable.
\section{General concept}\label{section1}
The Time-Gated Electric field Reconstruction (TIGER) technique relies on two basic principles: the creation of four identical temporally delayed sub-pulses from the pulse to be measured and the record of a second-order nonlinear optical effect with a camera. These two steps are respectively performed with a four-faced pyramid-like optical element and either by using the nonlinear response of a CCD camera or by imaging the second harmonic signal generated in a nonlinear crystal on the camera. The combination of these two elements makes the TIGER extremely compact and particularly easy to align. Let us first present how the four-faced pyramid-like optical element allows to create four time-delayed replicas.
\subsection{Geometrical considerations}
\begin{figure}
	 \includegraphics[width=\linewidth,keepaspectratio]{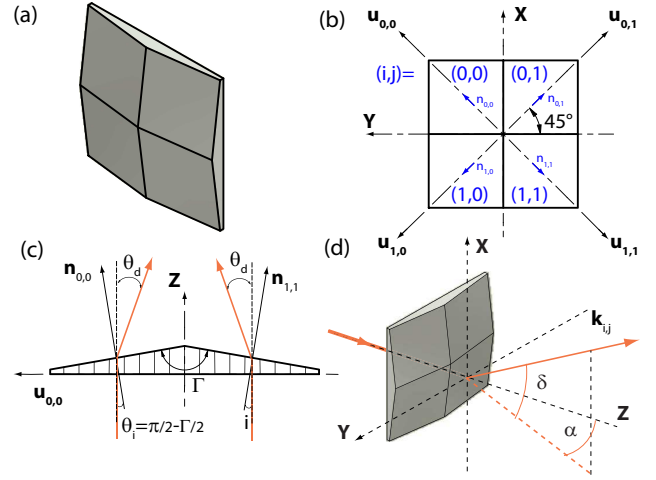}
	\caption{Four-faced pyramid-like optical element for creating four time-delayed replicas (a). Geometrical drawing of the pyramid (b,c) and deviation angles induced by the refraction of the input beam on one face of the pyramid (d). The beam arrives with an incidence angle $\theta_i=\pi/2-\Gamma/2$ and is deviated by an angle $\theta_d$, where $\Gamma$ is the diagonal apex of the pyramid.}
	\label{Figure1}
\end{figure}

The TIGER is based on recording a nonlinear effect induced by four time-delayed replicas. In order to probe all delays in a single shot, a four-faced pyramid-like optical element [Fig.\,\ref{Figure1}(a)], whose geometry is defined by the diagonal apex angle $\Gamma$ [see Figs.\,\ref{Figure1}(b-c)], is introduced in the input beam path.
%The side angle $\beta$ is linked to $\Gamma$ as
%\begin{equation}
%\beta=\textrm{atan}\left[\sqrt{\frac{1+\textrm{cos}\Gamma}{2\left(1-\textrm{cos}\Gamma\right)}}\right].
%\end{equation}
The input face of the pyramid is placed perpendicularly to the beam propagation axis $\mathbf{Z}$. The laser beam is sufficiently wide for illuminating the four faces of the pyramid. For each output face, defined by the indexes $\textrm{(i=0,1;j=0,1)}$ [see Fig.\,\ref{Figure1}(b)], the vector normal to the interface $\mathbf{n}_{\textrm{i,j}}$ is located in the diagonal plane and is given by
\begin{equation}
\mathbf{n}_{\textrm{i,j}}=\textrm{cos}\left(\Gamma/2\right)\,\mathbf{u_{\textrm{i,j}}}+\textrm{sin}\left(\Gamma/2\right)\,\mathbf{z},
\end{equation}
where $\mathbf{u_{\textrm{i,j}}}=\left[(-1)^\textrm{i}\,\mathbf{X}+(-1)^\textrm{j}\,\mathbf{Y}\right]/\sqrt{2}$ [see Figs.\,\ref{Figure1}(b-c)].
%\begin{equation}
%\vec{n}_{(i,j)}=\frac{1}{\sqrt{1+2\textrm{tan}^2\beta}}
% \begin{pmatrix}
% \pm\textrm{tan}\beta \\
% \pm\textrm{tan}\beta \\
% 1
% \end{pmatrix}.
%\end{equation}
After propagation through the pyramid, the input beam is split into four replicas. The output propagation directions of the replicas $\mathbf{k_{\textrm{i,j}}}$ are given by
\begin{equation}
\mathbf{k_{\textrm{i,j}}}=-\textrm{sin}\theta_d\,\mathbf{u_{\textrm{i,j}}}+\textrm{cos}\theta_d\,\mathbf{Z}=
 \begin{pmatrix}
\frac{(-1)^\textrm{i+1}}{\sqrt{2}}\textrm{sin}\theta_d \\
\frac{(-1)^\textrm{j+1}}{\sqrt{2}}\textrm{sin}\theta_d \\
\textrm{cos}\theta_d
 \end{pmatrix},
\end{equation}
where $\theta_d$ is the deviation angle that writes
\begin{equation}
\theta_d=\textrm{asin}\left[n\,\textrm{sin}\left(\frac{\pi-\Gamma}{2}\right)\right]-\frac{\pi-\Gamma}{2},
\end{equation}
%\theta_r=\textrm{asin}\left(\frac{n\sqrt{2}\textrm{tan}\beta}{\sqrt{1+2\textrm{tan}^2\beta}}\right)-\textrm{asin}\left(\frac{\sqrt{2}\textrm{tan}\beta}{\sqrt{1+2\textrm{tan}^2\beta}}\right),
with $n$ defining the refractive index of the material composing the pyramid at the considered wavelength.
Equivalently, the replicas are deviated from the horizontal plane by an angle $\delta_{\textrm{i,j}}=(-1)^\textrm{i+1}\textrm{asin}\left(\frac{1}{\sqrt{2}}\textrm{sin}\theta_d\right)$ and by an angle $\alpha_{\textrm{i,j}}=(-1)^\textrm{j+1}\textrm{atan}\left(\frac{1}{\sqrt{2}}\textrm{tan}\theta_d\right)$ with respect to $\mathbf{Z}$ [see Fig.\,\ref{Figure1}(d)]. After some propagation distance, the four replicas spatially overlap. However, the time at which they arrive depends on the spatial position in the plane $\left(x,y\right)$.
Using one of the replicas as a time reference, the three other replicas will arrive with delays $\tau_{1..3}$ that write
\begin{align}
\begin{split}
\tau_1&=\frac{\sqrt{2}\textrm{sin}\theta_d}{c}x,\\
\tau_2&=\frac{\sqrt{2}\textrm{sin}\theta_d}{c}y,\\
\tau_3&=\frac{\sqrt{2}\textrm{sin}\theta_d}{c}\left(x+y\right),
\end{split}
\label{Eq5}
\end{align}
where $c$ is the light velocity. Looking at Eqs.\,\ref{Eq5}, one immediately notices that $\tau_1$ (resp. $\tau_2$) only depends on the $x$ (resp. y) coordinate while $\tau_3=\tau_1+\tau_2$. It implies that, at the propagation distance where the four beams overlap, there is direct linear correspondence between the delays $(\tau_1,\tau_2)$ and the $(x,y)$ coordinates. The four-faced pyramid can then be viewed as a bi-dimensional extension of the Fresnel biprism, which is commonly used in single shot characterization devices for creating two time-delayed replicas in a single spatial dimension \cite{12,13}.
\begin{figure*}[t!]
	 \includegraphics[width=17cm,keepaspectratio]{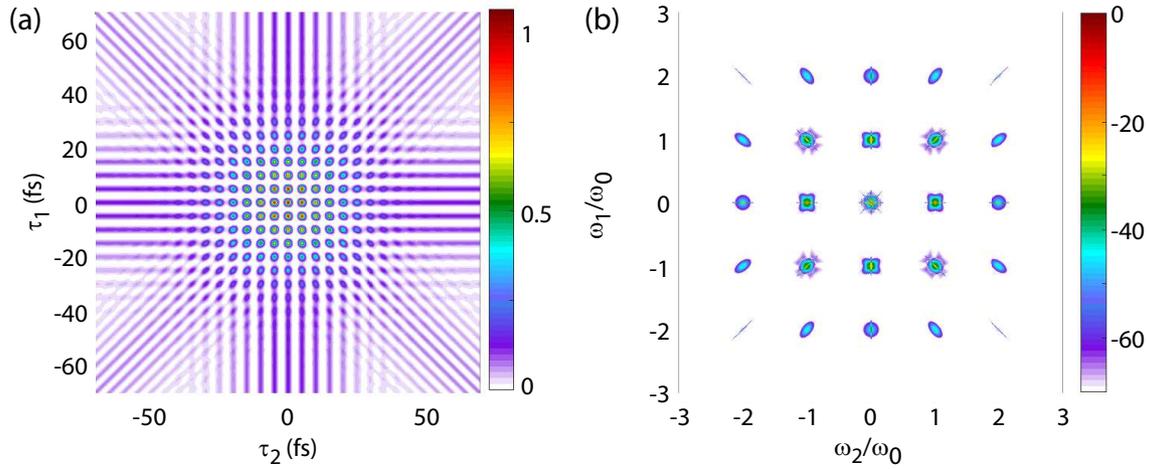}
	\caption{Theoretical TIGER signal calculated with Eq.\,\ref{Eq.Gen} for a 1.5\,$\mu$m Gaussian 30\,fs FTL laser pulse (a) and its two-dimensional Fourier transform (b).}
	\label{Figure2}
\end{figure*}
\subsection{Interferometric TIGER signal}
%figure*
Following the results obtained above, at the plane where the four replicas are spatially superimposed, the total electric field can be written as
$E(x,y,t,\tau_1,\tau_2)=\varepsilon_1(x,y,t)+\varepsilon_2(x,y,t-\tau_1)+\varepsilon_3(x,y,t-\tau_2)+\varepsilon_4(x,y,t-\tau_1-\tau_2)+c.c.$, where $\varepsilon_k(t)=A_k(x,y,t)\textrm{e}^{-i\omega_0t}$ is the complex field at central frequency $\omega_0$ coming from the $k^{\textrm{th}}$ face of the pyramid. For the following, it will be assumed that the four beams are perfectly balanced and that the field is sufficiently homogeneous in space for neglecting its spatial variation, i.e., $A_1(x,y,t)=A_2(x,y,t)=A_3(x,y,t)=A_4(x,y,t)=A(t)$. Using either second harmonic generation in a nonlinear crystal or two-photon absorption in a camera, the signal $S(\tau_1,\tau_2)$ reads
\begin{equation}
S(\tau_1,\tau_2)\propto\int{|E^2(t,\tau_1,\tau_2)|^2dt}.
\label{Eq.Gen}
\end{equation}
Figure\,\ref{Figure2}(a) shows a typical TIGER signal numerically calculated with Eq.\,\ref{Eq.Gen} for a 1.5\,$\mu$m Gaussian 30\,fs Fourier-limited (FTL) laser pulse.
After cumbersome but straightforward calculations, one obtains that $S$ can be written as the sum of 25 bi-dimensional distributions oscillating around $\left(m\omega_0,n\omega_0\right)$, with $[m,n]=0,\pm1,\pm2$:
\begin{equation}
S(\tau_1,\tau_2)\propto\sum_{m,n=0,\pm1,\pm2}{S_{m,n}(\tau_1,\tau_2)},
\label{Eq.Fourier}
\end{equation}
with $S_{m,n}(\tau_1,\tau_2)=F_{m,n}(\tau_1,\tau_2)e^{-i\omega_0\left(m\tau_1+n\tau_2\right)}$, where $F_{m,n}$ are slowly varying two-dimensional functions. Accordingly, the two-dimensional Fourier transform of $S$, $\widetilde{S}(\omega_1,\omega_2)$, embeds 25 distinct contributions as it can be seen in Fig.\,\ref{Figure2}(b).
Moreover, one can point out that $S(\tau_1,\tau_2)$ is a real-valued even function (with respect to both $\tau_1$ and $\tau_2$) and that the two variables $\tau_1$ and $\tau_2$ can be permuted. These properties indicates that the knowledge of only 6 contributions over the 25 (namely $S_{0,0},S_{1,0},S_{1,1},S_{2,0},S_{2,1}$, and $S_{2,2}$) is sufficient for recovering the full signal. After recording the full TIGER signal, each of these components can be isolated by an appropriate bi-dimensional spectral filtering. It is interesting to detail the analytical formula for some of the contributions. For instance, one has:
\begin{eqnarray}
\begin{aligned}
&F_{2,0}(\tau_1,\tau_2)=2\int{A^2(t)A^2(t-\tau_1)dt}&\\
&+4\int{A(t)A^*(t-\tau_1)A(t-\tau_2)A^*(t-\tau_1-\tau_2)dt},&
\end{aligned}
\end{eqnarray}
\begin{eqnarray}
\begin{aligned}
&F_{2,1}(\tau_1,\tau_2)=2\int{A^2(t)A^*(t-\tau_1)A^*(t-\tau_1-\tau_2)dt}&\\
&+2\int{A(t)A(t-\tau_2)A^{*2}(t-\tau_1-\tau_2)dt},&
\end{aligned}
\end{eqnarray}
\begin{equation}
\begin{aligned}
&F_{2,2}(\tau_1,\tau_2)=\int{A^2(t)A^{*2}(t-\tau_1-\tau_2)dt}.&
\end{aligned}
\end{equation}
So as to link the TIGER signal with the one obtained with SHG-FROG technique, let us express the Fourier transforms of $S_{2,0}$ and $S_{2,2}$ along the first dimension:
\begin{eqnarray}
\begin{aligned}
&\widetilde{S}_{2,0}(\omega_1,\tau_2)=4\left|\int{A(t)A(t-\tau_2)e^{i\left(\omega_1-2\omega_0\right)t}dt}\right|^2&\\
&+2\left|\int{A^2(t)e^{i\left(\omega_1-2\omega_0\right)t}dt}\right|^2,&\\
&\widetilde{S}_{2,2}(\omega_1,\tau_2)=\left|\int{A^2(t)e^{i\left(\omega_1-2\omega_0\right)t}dt}\right|^2e^{-i\omega_1\tau_2}.&
\end{aligned}
\end{eqnarray}
As a consequence, one has
\begin{equation}
\boxed{
\widetilde{S}_{2,0}-2\left|\widetilde{S}_{2,2}\right|\propto\left|\int{A(t)A(t-\tau_2)e^{i\left(\omega_1-2\omega_0\right)t}dt}\right|^2,
}
\label{FROG}
\end{equation}
which is nothing but the expression of the well-known SHG-FROG signal. The fact that the SHG-FROG signal is embedded in the TIGER signal indicates that this technique is able to fully characterize an ultrashort laser pulse. Finally, one can also express the Fourier transform of $S_{2,1}$ along the first dimension:
\begin{equation}
\widetilde{S}_{2,1}(\omega_1,\tau_2)=4e^{-i\frac{\omega_1\tau_2}{2}}\mathcal{R}e\left[\widetilde{\kappa}(\omega_1,\tau_2)e^{i(\omega_0-\frac{\omega_1}{2})\tau_2}\right],
\end{equation}
with
\begin{eqnarray}
\begin{aligned}
&\widetilde{\kappa}(\omega_1,\tau_2)=&\\
&\left(\int{A^2(t)e^{i(\omega_1-2\omega_0)t}dt}\right)^*\int{A(u)A(u-\tau_2)e^{i(\omega_1-2\omega_0)u}du}.&
\end{aligned}
\end{eqnarray}
\begin{figure}[t!]
	 \includegraphics[width=8.5cm,keepaspectratio]{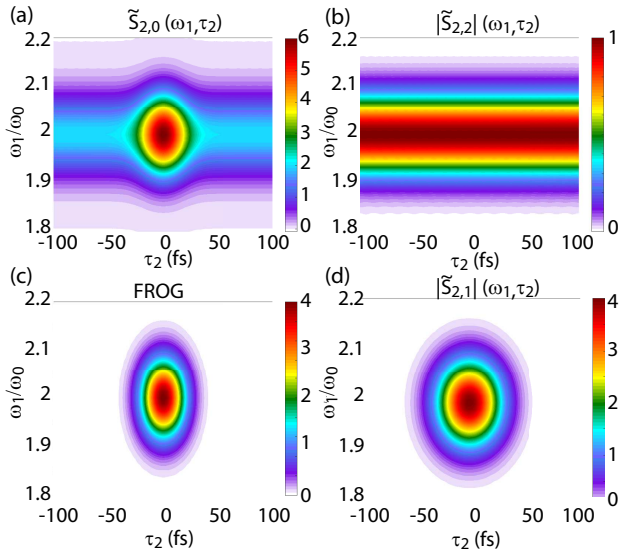}
	\caption{Contribution $\widetilde{S}_{2,0}$ (a) and $\widetilde{S}_{2,2}$ (b) obtained after a two-dimensional Fourier filtering of the full interferometric TIGER signal. Background-free Frequency-Resolved Optical Gating measurement obtained from (a,b) using Eq.\,\ref{FROG}. Contribution $\widetilde{S}_{2,1}$ (d) obtained after a two-dimensional Fourier filtering of the full interferometric TIGER signal.}
	\label{Figure3}
\end{figure}%
Figures\,\ref{Figure3}(a,b) show the contributions $\widetilde{S}_{2,0}$ and $\widetilde{S}_{2,2}$ extracted from the full TIGER signal displayed in Fig.\,\ref{Figure2}(a) by an appropriate two-dimensional spectral filtering. Figure\,\ref{Figure3}(c) displays the associated SHG-FROG signal obtained from Figs.\,\ref{Figure3}(a,b) using Eq.\,\ref{FROG}. Finally, Fig.\,\ref{Figure3}(d) displays the absolute value of $\widetilde{S}_{2,1}$ also retrieved from Fig.\,\ref{Figure2}(a). To conclude, the interferometric TIGER signal embeds all necessary information for fully characterizing an ultrashort laser pulse.\

\begin{figure}[t!]
	 \includegraphics[width=8.5cm,keepaspectratio]{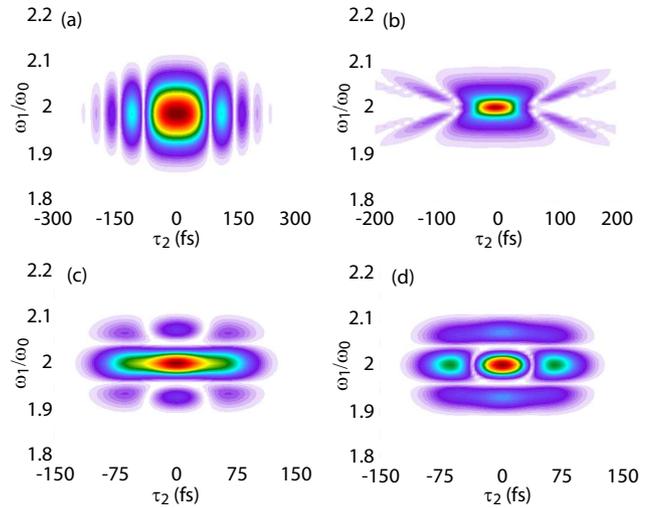}
	\caption{Modulus of the contribution $\widetilde{S}_{2,1}$ for a 30\,fs laser pulse with 1000\,fs$^2$ group delay dispersion (a) and with 50000\,fs$^3$ third-order dispersion (b). Contribution $\widetilde{S}_{2,1}$ for two 30\,fs in-phase (c) and out-of-phase (d) pulses separated by 60\,fs.}
	\label{Figure4-5}
\end{figure}

\begin{figure}[h!]
	 \includegraphics[width=8.5cm,keepaspectratio]{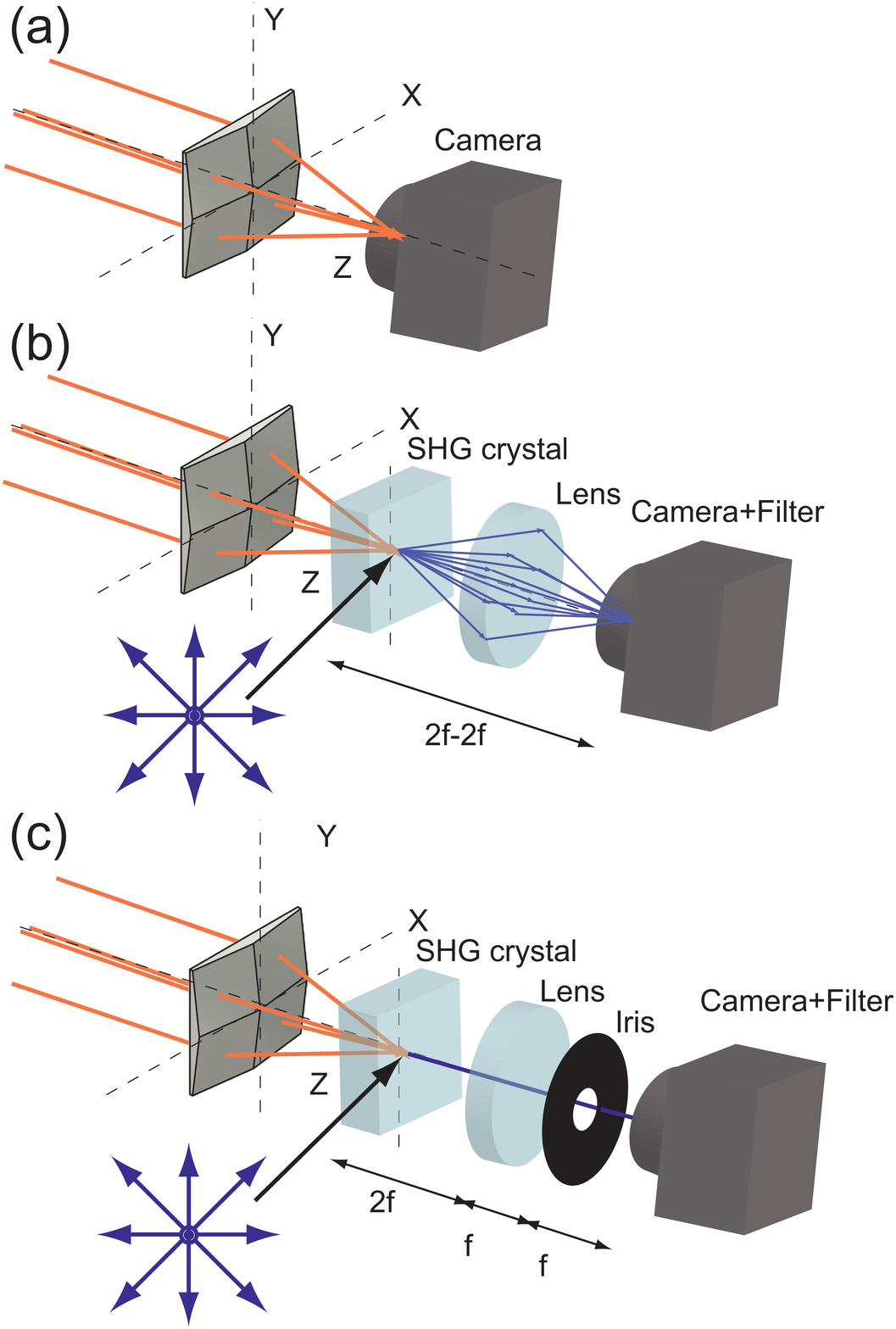}
	\caption{Two-photon absorption (a) and SHG (b) based interferometric TIGER apparatus. Intensimetric TIGER signal using second harmonic generation (c). In this case, an iris diaphragm is placed in the Fourier plane of the imaging lens so as to select only the second harmonic radiation propagating along the $z$ direction.}
	\label{Figure4}
\end{figure}

So as to retrieve the pulse amplitude and phase, one has to use an iterative algorithm so as to fit the map signal. However, considering that the full TIGER interferometric pattern is not mandatory for characterizing a pulse, one can choose to retrieve the pulse characteristics from one specific component of the signal. In particular, one can choose to use the contribution $\widetilde{S}_{2,0}$ (namely the one containing the FROG information), but it appears that this contribution is not background-free (along $\tau_2$) which is a main drawback. The background theoretically corresponds to twice the absolute value of the $\widetilde{S}_{2,2}$ contribution so that one could be tempted to extract it by subtracting from $\widetilde{S}_{2,0}$ the contribution calculated from $\widetilde{S}_{2,2}$ (as shown in Eq.~\ref{FROG}). However, from an experimental point of view, since measurements inevitably include noise that differs for the two contributions, this strategy appears not to be very convenient. As a consequence, we chose to use the contribution $\widetilde{S}_{2,1}$ in order to retrieve the temporal and spectral characteristics of the pulse. This contribution has the advantage to be inherently background-free. Figure\,\ref{Figure4-5} displays the signal $\widetilde{S}_{2,1}$ for different spectral phase [Figs.\,\ref{Figure4-5}(a,b)] or amplitude [Figs.\,\ref{Figure4-5}(c,d)] modulations applied to a 30\,fs laser pulse. The clear phase and amplitude sensitivity of the obtained pattern clearly confirms that this contribution can be used for the retrieval procedure.
Figure\,\ref{Figure4}(a) depicts the typical experimental setup for measuring the interferometric TIGER signal in the case where direct two-photon absorption in a camera can be used. This is possible if the energy bandgap of the semiconductor is comprised between one and two times the photon energy. For instance, with a silicon camera, this is possible for laser central wavelengths comprised between around 1.4 and 2.4\,$\mu$m. Figure\,\ref{Figure4}(b) shows the typical experimental TIGER setup if second harmonic generation is used as a nonlinear effect. When the four replicas nonlinearly interact within the SHG crystal, second harmonic radiations are generated along nine distinct directions. The interferometric TIGER signal is then obtained by imaging the SHG-crystal with a lens that collect all the nine radiations and a dichroic filter placed before the camera rejects the fundamental radiation. However, as noticed just before, recording the full interferometric TIGER signal is not necessary for the complete temporal characterization of a pulse. In the next section, an all-optical method for obtaining an intensimetric TIGER signal will be presented in the case where second harmonic generation is used as a nonlinear process.
\subsection{Intensimetric TIGER signal in case of second harmonic generation}\label{SectionIntensimetric}
As explained in the previous section, using second harmonic generation as a nonlinear process leads to generate second harmonic radiations propagating along nine distinct directions. This can be easily understood by analyzing the direction of the wavevectors of the four pump pulse replicas. Each second harmonic radiation, propagating along a particular direction, can then be easily optically isolated by placing an iris diaphragm in the Fourier plane of the imaging lens [see Fig.\,\ref{Figure4}(c)]. In particular, a part of second harmonic radiation is generated along the $\mathbf{Z}$ axis direction. The signal $S$ induced by this contribution alone (i.e., after filtering the eight other contributions in the Fourier plane) recorded with a camera in the image plane of the imaging lens reads
\begin{equation}
S(\tau_1,\tau_2)\propto\int{\left|A(t)A(t-\tau_1-\tau_2)+A(t-\tau_1)A(t-\tau_2)\right|^2dt}.
\label{EqTIGERintensimetric}
\end{equation}
\begin{figure*}[h!]
	 \includegraphics[width=17cm,keepaspectratio]{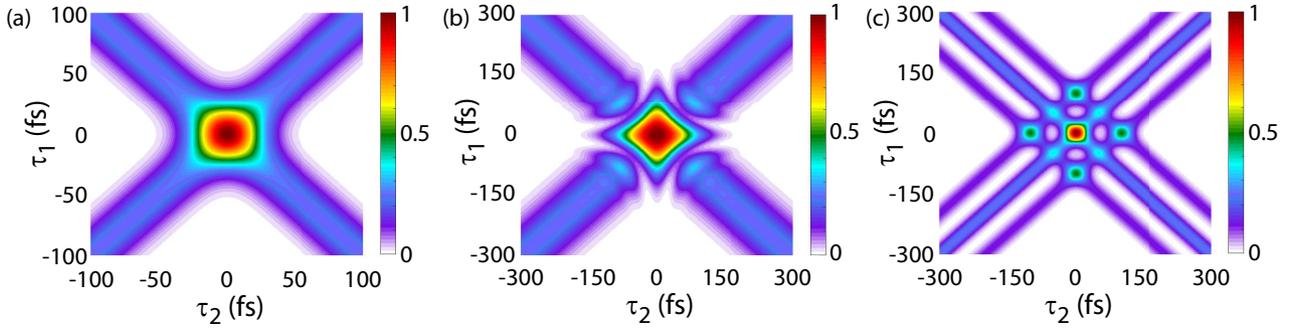}
	\caption{Intensimetric TIGER signal for a 30\,fs Fourier-transform limited pulse (a), a 30\,fs pulse chirped to 100\,fs (b), and two 30\,fs Fourier-transform limited pulses separated by 100\,fs (c).}
	\label{Figure5}
\end{figure*}
According to Eq.\,\ref{EqTIGERintensimetric}, the signal recorded this way is intensimetric, i.e., the fast oscillating components within the TIGER signal vanish. One can also notice that the signal embeds a background signal occurring along the two diagonals defined by $\tau_1=\tau_2$ and $\tau_1=-\tau_2$. Moreover, for $\tau_1=0$ (or $\tau_2=0)$, the intensimetric TIGER signal is proportional to the second-order pulse autocorrelation trace.
Figure\,\ref{Figure5} shows three different examples of intensimetric TIGER signal. Figure\,\ref{Figure5}(a) corresponds to the signal calculated for a 30\,fs [at full width at half-maximum (FWHM)] FTL pulse, while Fig.\,\ref{Figure5}(b) corresponds to the one calculated for the same pulse but chirped to 100\,fs (FWHM). Finally, Fig.\,\ref{Figure5}(c) corresponds to the TIGER signal obtained for two 30\,fs FTL pulses separated by 100\,fs. The three TIGER maps show noticeable modifications. One has to note that the TIGER exhibits the same indeterminacy as all FROG devices based on a second-order nonlinear process. For instance, the TIGER signal cannot determine the sign of the spectral phase since a negatively chirped pulse provides exactly the same signal as a positively chirped one. In addition, unlike the interferometric method where the fringes give access to the absolute carrier frequency, one cannot determine the central frequency of the laser pulse in case of intensimetric measurement. Intensimetric SHG-TIGER has nevertheless the advantage to release optical constraints imposed by the need to resolve the optical fringes produced by the interferometric method.
\section{Results}\label{section2}
In this section, the TIGER devices and experimental pulse characterizations for two different configuration are detailed. First, interferometric measurements obtained for a 70\,fs noncollinear optical parametric amplifier (NOPA) pumped by a 100\,fs Ti:Sa femtosecond laser and emitting between 1.4\,$\mu$m and 2.4\,$\mu$m are presented. Then, intensimetric SHG-TIGER measurements, recorded with a 35\,fs 800\,nm Ti:Sa femtosecond laser, are discussed.
\subsection{Interferometric two-photon absorption measurements}
\begin{figure}[!]
	 \includegraphics[width=8.5cm,keepaspectratio]{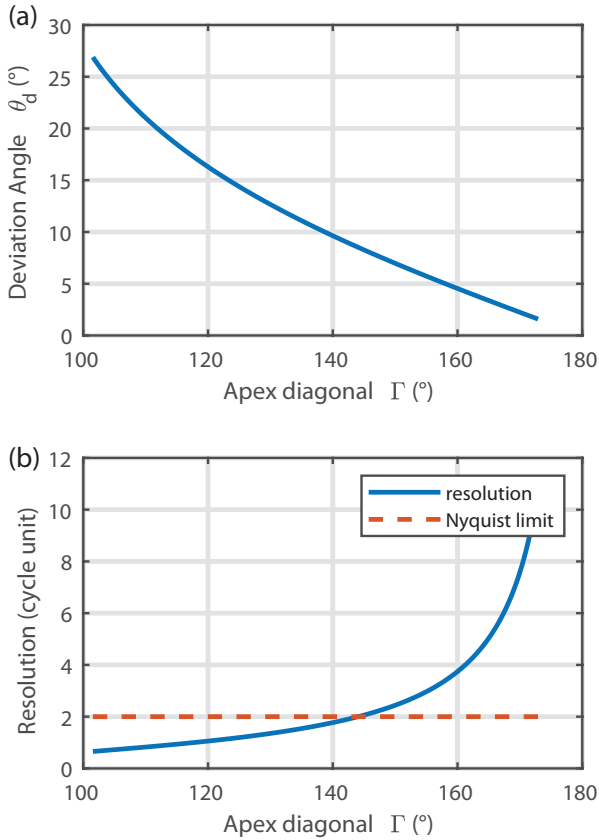}
	\caption{Deviation angle $\theta_d$ at $1.4\,\mu$m as a function of the diagonal apex of the pyramid $\Gamma$ (a) and related resolution for a pixel size $dx$=1.67\,$\mu$m (b).}
	\label{Figure6}
\end{figure}
\subsubsection{Experimental setup}
Since the central wavelength of the pulses emitted by the NOPA is between 1.4 and 2.4\,$\mu$m, two-photon absorption signal in a silicon (bandgap $\simeq$1.1\,eV) camera can be used for recording the TIGER signal. Recording an interferometric TIGER signal implies a spatial resolution of the fast oscillations present within the signal. According to above calculations, the fastest oscillations within the signal have a carrier frequency located around 2$\omega_0$. From Eqs.\,\ref{Eq5} and following the Nyquist theorem, this implies
\begin{equation}
\frac{\sqrt{2}\textrm{sin}\theta_r}{c}dx<\frac{\pi}{2\omega_0},
\end{equation}
where $dx$ is the pixel size of the camera. For this experiment, we used a 12 bit CMOS camera with pixel size $dx$=1.67\,$\mu$m. Figure\,\ref{Figure6}(a) shows the deviation angle $\theta_d$ as a function of the diagonal apex of the pyramid (fused silica) at a central wavelength $1.4\,\mu$m, which is the lowest central wavelength that can be measured by two-absorption in a silicon camera. Considering the pixel size $dx$=1.67\,$\mu$m, this then means that the diagonal apex of the pyramid has to be higher than about 145$^{\circ}$ [see Fig.\,\ref{Figure6}(b)]. We finally opted for $\Gamma=160^{\circ}$ for the experiment.
\subsubsection{Experimental results}
\begin{figure}[!]
	 \includegraphics[width=8.5cm,keepaspectratio]{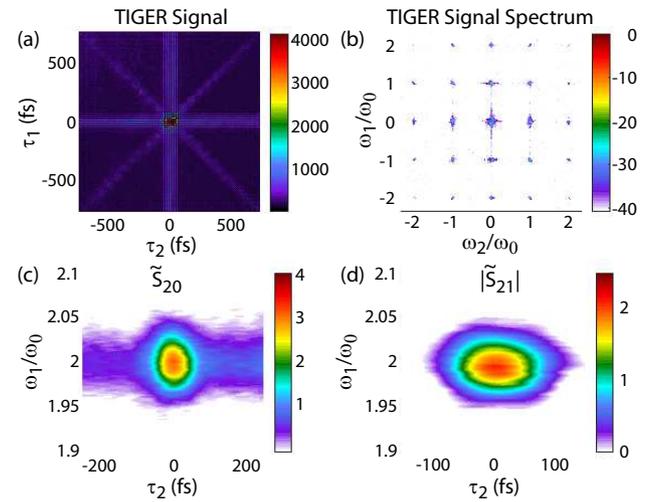}
	\caption{Experimental two-photon absorption based TIGER signal recorded at $\lambda_0$=1.8\,$\mu$m (a) and its two-dimensional Fourier transform (b). Signal $\widetilde{S}_{2,0}$ (c) and $\widetilde{S}_{2,1}$ (d) extracted from the TIGER signal after spectral filtering.}
	\label{Figure7}
\end{figure}
The first experiment was devoted to the characterization of the pulses directly coming from the NOPA. The central wavelength was set to $\lambda_0$=1.8\,$\mu$m. An energy of approximatively 1\,$\mu$J is used for performing single shot measurements. Figure \ref{Figure7}(a) shows a typical interferometric signal recorded in this case. The corresponding bi-dimensional Fourier transform is depicted in Fig.\,\ref{Figure7}(b). After an adequate spectral filtering, the corresponding contributions $\widetilde{S}_{2,0}$ and $\widetilde{S}_{2,1}$ can be isolated [see Figs.\,\ref{Figure7}(c,d)].
\begin{figure}[h!]
	 \includegraphics[width=8.5cm,keepaspectratio]{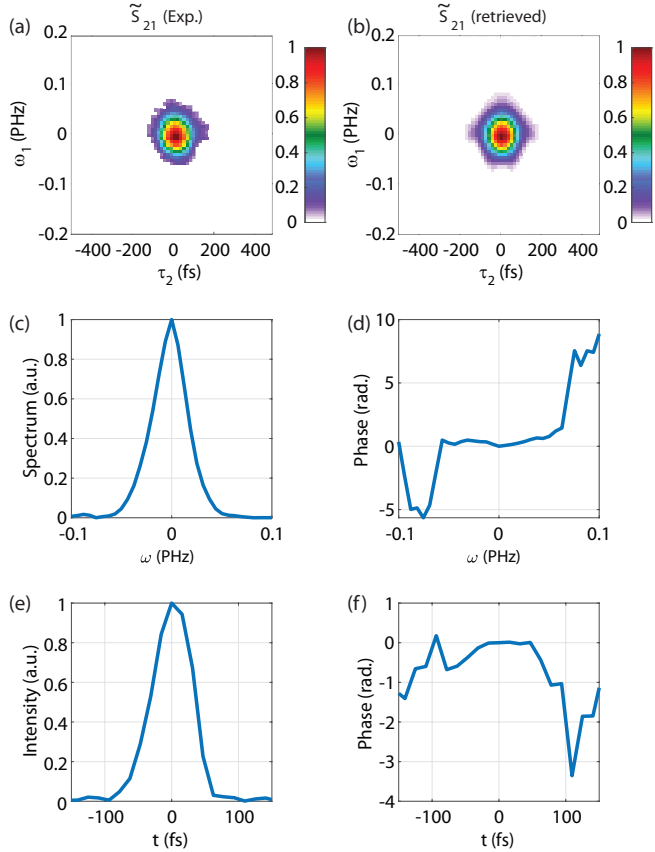}
	\caption{Experimental signal $\widetilde{S}_{2,1}(\omega_1,\tau_2)$ (a) and retrieved signal (b) for the laser pulse ($\lambda_0=\,$1.8$\,\mu$m) coming from the NOPA. Retrieved spectral intensity (c) and phase (d). Retrieved temporal intensity profile (e) and phase (f).}
	\label{Figure8}
\end{figure}
In order to retrieve the spectral phase and amplitude of the laser pulse, we use the contribution $S_{2,1}$, which is background free and therefore less sensitive to the spatial noise. For the retrieval procedure, the signal $\tilde{S}_{2,1}(\omega_1,\tau_2)$ is then undersampled on a 64x64 frequency-temporal grid and a Levenberg-Marquardt-based algorithm \cite{14} is used for retrieving both amplitude and phase of the field in the spectral domain.
\begin{figure}[h!]
	 \includegraphics[width=8.5cm,keepaspectratio]{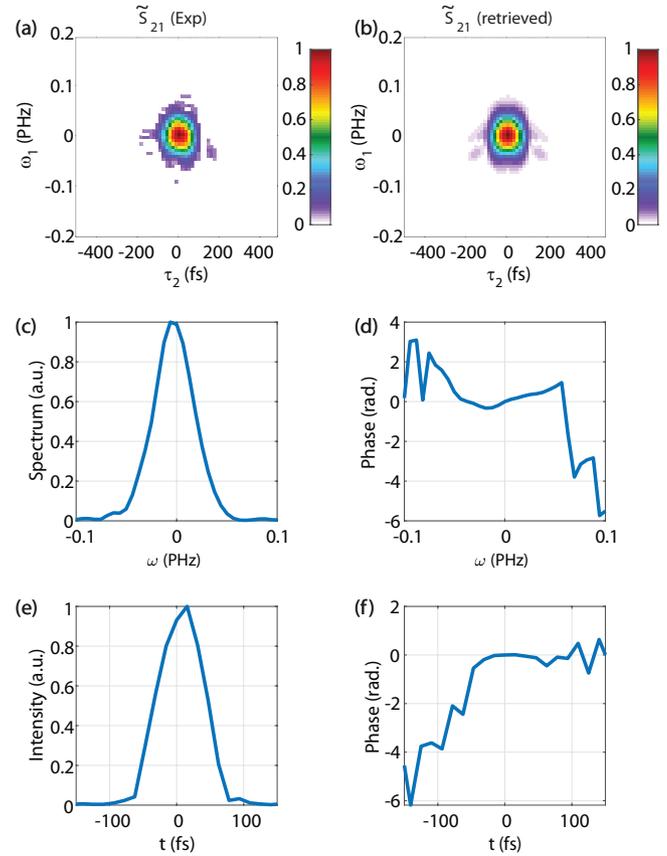}
	\caption{Experimental signal $\widetilde{S}_{2,1}(\omega_1,\tau_2)$ (a) and retrieved signal (b) for the laser pulse ($\lambda_0=\,$1.8$\mu$m) after propagation within a 1.8\,cm fused silica plate. Retrieved spectral intensity (c) and phase (d). Retrieved temporal intensity profile (e) and phase (f).}
	\label{Figure9}
\end{figure}
The results of the retrieval by the algorithm is displayed in Fig.\,\ref{Figure8}. The retrieved pulse has a duration (at full-width at half-maximum) of approximately 70\,fs with a small residual group delay dispersion (around $\pm$370\,fs$^2$). In order to test our device, a 1.8\,cm thick fused silica plate is inserted before the TIGER.
The results of the retrieval procedure are shown in Fig.\,\ref{Figure9}. In this case, the pulse is found to be slightly longer (about 78\,fs) with a residual group delay dispersion of $\pm$872\,fs$^2$. The group delay dispersion introduced by the plate is then evaluated to be approximately 1240\,fs$^2$ (in absolute value), close the theoretical group delay dispersion of fused silica (-1130\,fs$^2$) at the considered wavelength. This observation demonstrates the sensitivity of the TIGER method with respect to the spectral phase of the pulse. In order to also check its sensitivity with respect to the spectral amplitude adequately, a sequence of two FTL pulses separated by 140\,fs is produced by using a calibrated multiple order waveplate and a polarizer.
Figure\,\ref{Figure10} shows the obtained results. The TIGER successfully reconstructs the field amplitude spectrum, which is characterized by a sinusoidal modulation of the amplitude along with a sequence of $\pi$-steps for the spectral phase. In the time domain, the retrieval process correctly reproduces the two pulses separated by 140\,fs.

\subsection{Intensimetric second harmonic generation measurements}
\begin{figure}
	 \includegraphics[width=8.5cm,keepaspectratio]{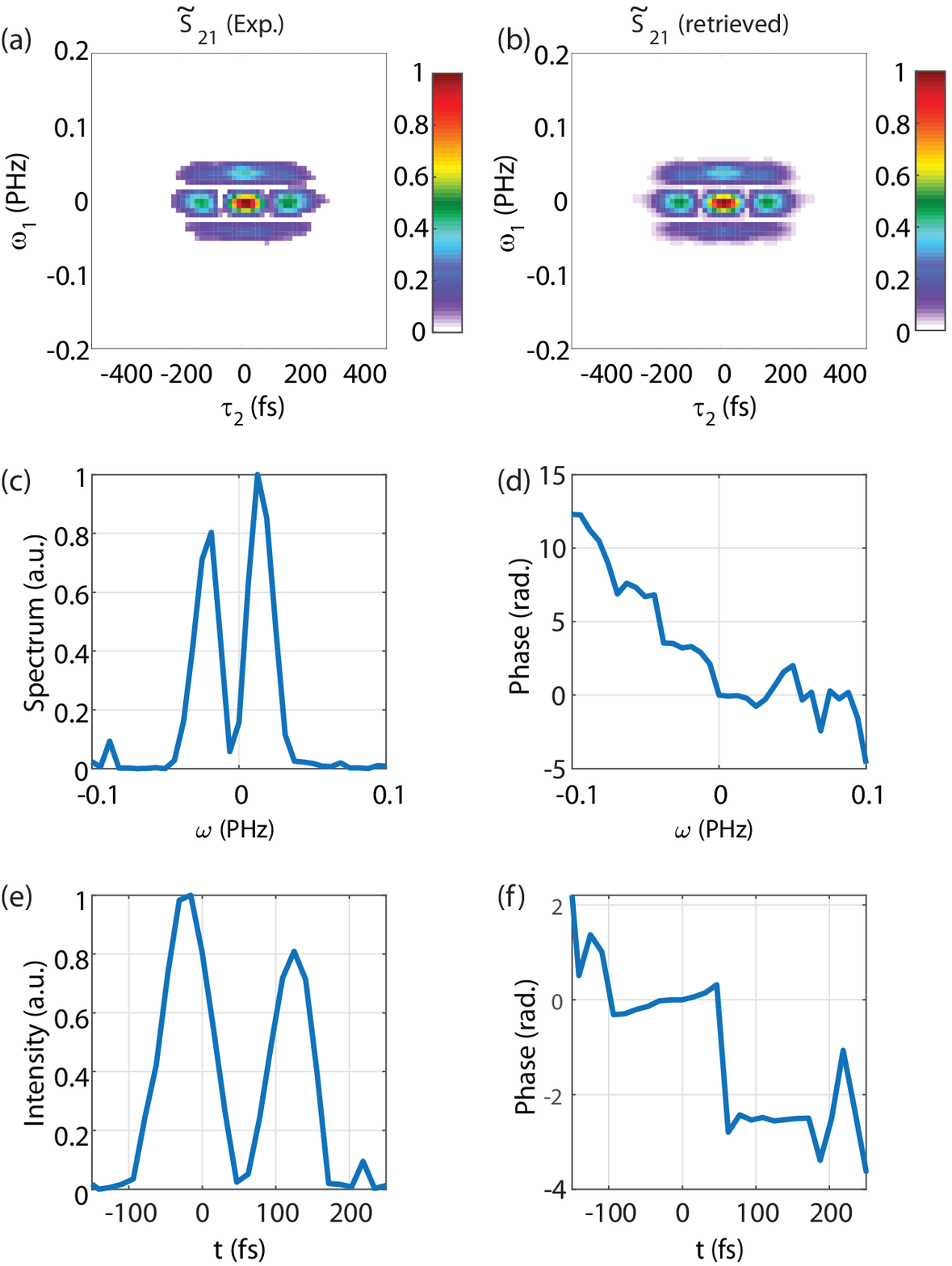}
	\caption{Experimental signal $\widetilde{S}_{2,1}(\omega_1,\tau_2)$ (a) and retrieved signal (b) for the laser pulse ($\lambda_0=\,$1.8$\mu$m) for the 140\,fs-separated 70\,fs bipulse. Retrieved spectral intensity (c) and phase (d). Retrieved temporal intensity profile (e) and phase (f).}
	\label{Figure10}
\end{figure}
Despite its extreme simplicity, the two-photon absorption-based interferometric TIGER is limited to a given spectral range depending on the material composing the pixel of the camera. For central wavelength ranging from 1.4 to 2.4\,$\mu$m, a simple silicon camera can be used, as shown in the above section. For central wavelengths ranging from approximatively 1.7 to 3.4\,$\mu$m, a camera with InGaAs based pixels could be used. If one wants to extend the concept to lower wavelengths, to the best of our knowledge, no bi-dimensional sensor with sufficiently small pixels and allowing two-photon absorption is yet so far available. In order to generalize the concept for lower wavelengths, a second harmonic generation based TIGER was developed. As explained in section\,\ref{SectionIntensimetric}, recording the full interferometric signal is not mandatory for retrieving the temporal and spectral characteristics of the pulse. The advantage of using an intensimetric approach in this case is to lift the necessity to spatially resolve the interference fringes, which can be somehow challenging when using an imagery optical system. This section describes the experimental setup and the  experimental results obtained with the second harmonic generation-based intensimetric TIGER setup.
\subsubsection{Experimental setup}
\begin{figure}
	 \includegraphics[width=8.5cm,keepaspectratio]{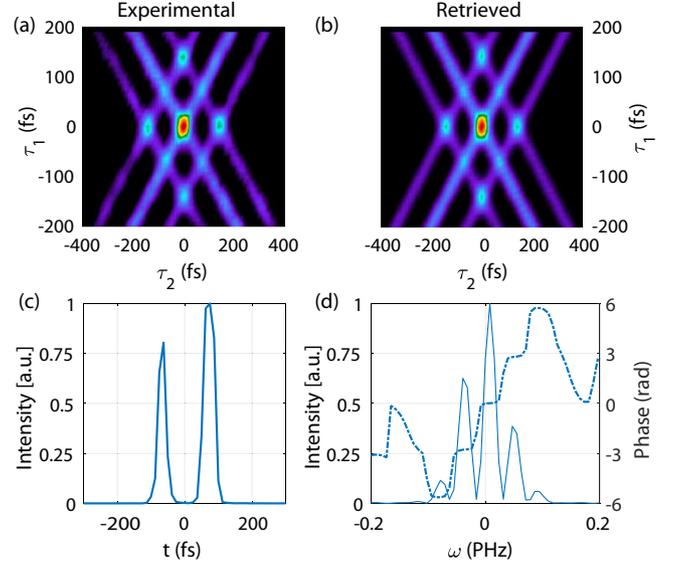}
	\caption{Experimental (a) and retrieved (b) SHG-based intensimetric TIGER signal in case of a 140\,fs-separated 35\,fs bipulse. Retrieved temporal intensity profile (c). Retrieved spectral intensity (solid blue) and phase (dashed-dotted line) (d).}
	\label{Figure11}
\end{figure}
Figure\,\ref{Figure4}(c) shows the principle of the experimental setup developed for characterizing our 35\,fs Ti:Sa laser pulse. The diagonal apex of the pyramid is $160^{\circ}$. The SHG crystal is a 50\,$\mu$m thick BBO. The imaging lens is a $f=$3\,cm focal length biconvex lens placed at a distance 2$f$. After the lens, an iris diaphragm is placed in the Fourier plane (i.e., at a distance $f$ after the lens) so as to select only the second harmonic radiation that propagates along the $z$ axis. Then, the camera is placed at 4$f$ from the crystal. In the case of intensimetric measurements, one has to calibrate the magnification of the imaging system, which was not the case with the interferometric two-photon absorption-based setup. So as to calibrate the magnification of the imaging system, a bipulse (two identical pulses separated by 140\,fs) was created by using a calibrated multi-order waveplate and a polarizer. In this case, three lobes separated by $\Delta\tau_1=$ 140\,fs are expected at $\tau_2=0$ (and vice versa). It then allows to find the space-time calibration coefficient for our setup. Figure\,\ref{Figure11}(a) shows the experimental TIGER signal obtained with this temporal shaping. The retrieved TIGER signal is displayed in Fig.\,\ref{Figure11}(b), while the temporal intensity profile (resp. spectral intensity and phase) is shown in Fig.\,\ref{Figure11}(c) [resp. Fig.\,\ref{Figure11}(d)]. As shown, the fitting procedure successfully retrieves the temporal and spectral characteristics of the laser pulse. Again, the present configuration successfully reconstructs the sinusoidal modulation of the amplitude along with the sequence of $\pi$-steps for the spectral phase demonstrating the sensitivity of the measurement with respect to both spectral phase and amplitude.

\subsubsection{Experimental results}
\begin{figure}
	 \includegraphics[width=8.5cm,keepaspectratio]{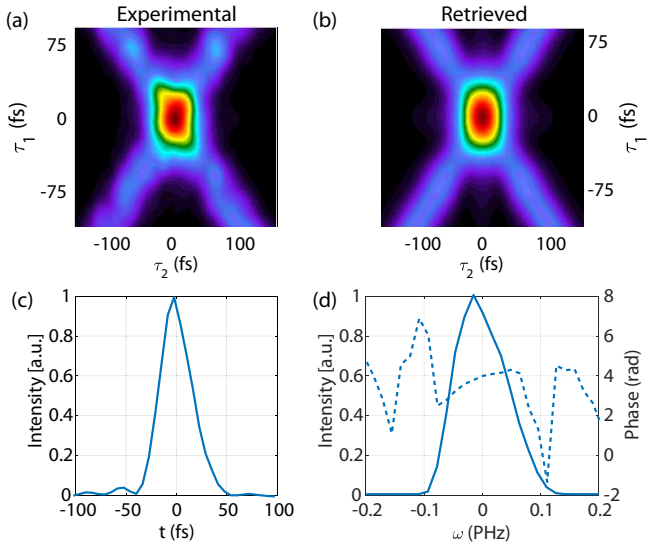}
	\caption{Experimental (a) and retrieved (b) SHG-based intensimetric TIGER signal for the laser output pulse. Retrieved temporal intensity profile (c). Retrieved spectral intensity (solid blue) and phase (dashed-dotted line) (d).}
	\label{Figure12}
\end{figure}

We then use the intensimetric TIGER device for characterizing the output laser pulse. An input pulse energy of approximatively 1.2\,$\mu$J was used for single shot operation. Figure\,\ref{Figure12} shows the final outputs of the measurements. The pulse duration is found to be approximately 34\,fs, close to the FTL limit. This observation is in good agreement with the relatively flat spectral phase observed in the retrieved spectral field in Figure\,\ref{Figure12}(d).

\begin{figure}
	 \includegraphics[width=8.5cm,keepaspectratio]{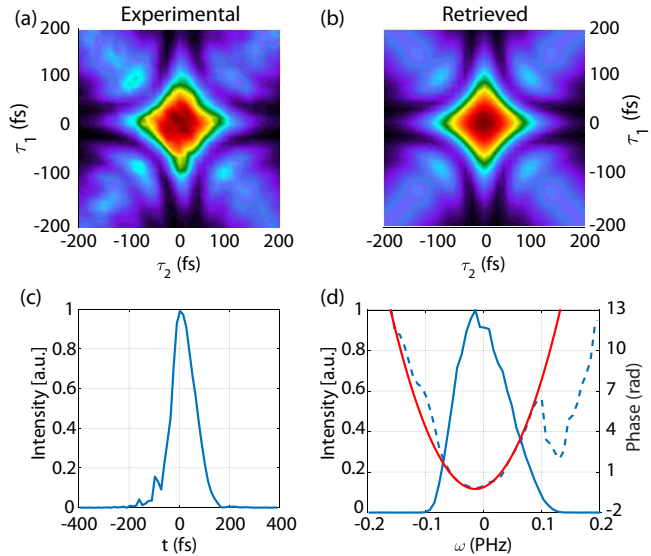}
	\caption{Experimental (a) and retrieved (b) SHG-based intensimetric TIGER signal after propagation through a 5\,mm SF11 plate. Retrieved temporal intensity profile (c). Retrieved spectral intensity (solid blue) and phase (dashed-dotted line) (d). The red solid line in (d) corresponds to the theoretical phase introduced by the SF11 plate.}
	\label{Figure13}
\end{figure}

Finally, another measurement has been done by putting in the laser beam path a 5\,mm thick SF11 plate, introducing a calibrated group delay dispersion of 950\,fs$^2$. The measurement and retrieval are shown in Fig.\,\ref{Figure13}. As shown, the phase retrieved by the TIGER is again in very good agreement with the expectation.
\section{Conclusion}
In this paper, an original device enabling a single shot and full characterization of ultrashort laser pulses is presented. Unlike all other existing techniques, the TIGER device does not use any spectral measurement. Instead, the device is based on recording a second-order nonlinear effect created by four time-delayed replicas of the pulses to be measured. These replicas are generated by inserting a four-faces pyramid like optical element that converts the two transverse directions in two independent time delays. After describing the general principle operation of the device, theoretical and experimental results using two-photon absorption performed directly in the camera sensor were presented. This approach, which is relevant for the characterization of infrared laser pulses, is limited in term of wavelength range. We have therefore shown (both theoretically and experimentally) that second harmonic generation can also be exploited. In this case, the signal can be simplified by filtering a part of the interferometric signal via a spatial Fourier filtering. The performance of the overall set-up for the pulse reconstruction is of high quality for both configurations. The device is extremely compact and easy to align. It does not require any spectrometer or imaging spectrometer and the intensimetric configuration only needs a relatively low resolution camera. Finally, it is worth to emphasize that the four-faces pyramid like optical element for generating four time-delayed pulses could also be used for performing single shot two-dimensional spectroscopy experiment.
\begin{acknowledgement}
This work has benefited from the facilities of the SMARTLIGHT platform in Bourgogne Franche-Comt\'e (EQUIPEX+ contract "ANR-21-ESRE-0040").
This work has been supported by the EIPHI Graduate School (contract ANR-17-EURE-0002), Bourgogne-Franche-Comt\'e Region, the CNRS, and the SATT Grand-Est. We thank the CRM-ICB (Brice Gourier and Julien Lopez) for the mechanical realization of the intensimetric device and the CRI-CCUB for CPU loan on the multiprocessor server.
\end{acknowledgement}
\bibliographystyle{lpr}

\end{document}